\documentclass[aps,superscriptaddress,superbib,twocolumn,groupedaddress]{revtex4}

\usepackage[highlight]{neel}

\bibpunct{[}{]}{,}{n}{}{}

\graphicspath{{fig/}}

\usepackage[latin1]{inputenc}
\usepackage[T1]{fontenc}

\def\figurename{Fig.}

\usepackage[paperwidth=8.5in, paperheight=11in]{geometry}
\usepackage{bbm} 

\usepackage{bm} 
\usepackage{graphics} 
\usepackage{times}
\usepackage{bbm}
\usepackage{tabularx}
\usepackage{sistyle}
\usepackage{lipsum}
\usepackage{url}
\usepackage{float}
\usepackage{pdfpages} 
\usepackage{fancyhdr} 

\begin{document}
\title{Bloch-point-mediated topological transformations of magnetic domain walls in cylindrical nanowires}
\author{A. Wartelle}
\email{alexis.wartelle@neel.cnrs.fr}
\affiliation{Univ. Grenoble Alpes, CNRS, Grenoble INP\footnote{Institute of Engineering Univ. Grenoble Alpes}, Institut N\'{e}el, F-38000 Grenoble, France}
\author{B. Trapp}
\affiliation{Univ. Grenoble Alpes, CNRS, Grenoble INP\footnote{Institute of Engineering Univ. Grenoble Alpes}, Institut N\'{e}el, F-38000 Grenoble, France}
\author{M. Sta\v{n}o}
\altaffiliation[Present address: ]{CEITEC - Central European Institute of Technology, Brno University of Technology, 612 00 Brno, Czech Republic}
\affiliation{Univ. Grenoble Alpes, CNRS, Grenoble INP\footnote{Institute of Engineering Univ. Grenoble Alpes}, Institut N\'{e}el, F-38000 Grenoble, France}
\author{C. Thirion}
\affiliation{Univ. Grenoble Alpes, CNRS, Grenoble INP\footnote{Institute of Engineering Univ. Grenoble Alpes}, Institut N\'{e}el, F-38000 Grenoble, France}
\author{S. Bochmann}
\affiliation{Friedrich-Alexander Universit\"{a}t Erlangen-N\"{u}rnberg, Erlangen 91058, Germany}
\author{J. Bachmann}
\affiliation{Friedrich-Alexander Universit\"{a}t Erlangen-N\"{u}rnberg, Erlangen 91058, Germany}
\affiliation{Saint-Petersburg State University, Institute of Chemistry, Universitetskii pr. 26, 198504 St. Petersburg, Russia}
\author{M. Foerster}
\affiliation{Alba Synchrotron Light Facility, CELLS, Barcelona, Spain}
\author{L.~Aballe}
\affiliation{Alba Synchrotron Light Facility, CELLS, Barcelona, Spain}
\author{T. O. Mente\c{s}}
\affiliation{Elettra-Sincrotrone Trieste, S.C.p.A., Trieste I-34012, Italy}
\author{A. Locatelli}
\affiliation{Elettra-Sincrotrone Trieste, S.C.p.A., Trieste I-34012, Italy}
\author{A. Sala}
\affiliation{Elettra-Sincrotrone Trieste, S.C.p.A., Trieste I-34012, Italy}
\author{L. Cagnon}
\affiliation{Univ. Grenoble Alpes, CNRS, Grenoble INP\footnote{Institute of Engineering Univ. Grenoble Alpes}, Institut N\'{e}el, F-38000 Grenoble, France}
\author{J.-C. Toussaint}
\affiliation{Univ. Grenoble Alpes, CNRS, Grenoble INP\footnote{Institute of Engineering Univ. Grenoble Alpes}, Institut N\'{e}el, F-38000 Grenoble, France}
\author{O. Fruchart}
\email{olivier.fruchart@cea.fr}
\affiliation{Univ. Grenoble Alpes, CNRS, CEA, Grenoble INP\footnote{Institute of Engineering Univ. Grenoble Alpes}, INAC-Spintec, F-38000 Grenoble, France}

\date{\today}

\begin{abstract}
  Cylindrical nanowires made of soft magnetic materials, in contrast to thin strips, may host domain walls of two distinct topologies. Unexpectedly, we evidence experimentally the dynamic transformation of topology upon wall motion above a field threshold. Micromagnetic simulations highlight the underlying precessional dynamics for one way of the transformation, involving the nucleation of a Bloch-point singularity, however, fail to reproduce the reverse process. This rare discrepancy between micromagnetic simulations and experiments raises fascinating questions in material and computer science.
\end{abstract}

\maketitle

 Directional orders, such as nematics and ferromagnets, may give rise to topologically non-trivial textures of the order parameter. In ferromagnets, the variety of exchange interactions and host systems translates into a broad spectrum of such textures, such as non-zero Chern numbers in band structures on a kagome lattice\cite{bib-TAI2006}, merons in coupled disks\cite{bib-PHA2012} or multilayers\cite{bib-HIE2017}, chiral domain walls (DWs)\cite{bib-CHE2015,bib-TET2015,bib-CHE2015c} and skyrmions\cite{bib-MOR2016,bib-HEI2011b,bib-BOU2016}. Yet, while all those have a continuous spin texture, a singular configuration was theoretically predicted in 1965\cite{bib-FEL1965}: the Bloch point (BP). This is a point defect for the unit magnetization vector field $\vectm$, and as such the only possible topological defect in ferromagnetism\cite{bib-BRA2012}. It is the analogue of defects seen in nematic liquid crystals\cite{bib-WIL1972,bib-SAU1973}, for which the distribution of the director around the defect covers the unit sphere $\mathcal{S}^2$ exactly once. For this reason, an integer called winding number\cite{bib-WHI1947} is associated to the BP, which is its signature as a topological defect.

 The existence of the BP is crucial, as simulation suggested that the transformation from one spin texture to another of different topology is mediated by a~BP expulsion or injection: in static\cite{bib-THI2003} or dynamic\cite{bib-VAN2006} magnetization switching of vortex cores in thin films, the nucleation of skyrmions in dots\cite{bib-SAM2013b} or of DWs in magnetically soft cylindrical nanowires \cite{bib-HER2002a}. The latter system appears as a textbook playground for the investigation of topological transformations and of BPs. Indeed, BPs should exist at rest, unlike those involved in the dynamical transformation processes mentioned above. In detail, two types of DWs were predicted to exist in cylindrical nanowires, with different topologies. First is the Bloch point domain wall (BPW, also called vortex wall by some), hosting a BP at its center even at rest. The BPW  was predicted to reach a steady-state motion with high axial velocity even at high magnetic field\cite{bib-THI2006,bib-HER2015,bib-HER2016}. Second is the transverse-vortex wall (TVW, also called transverse wall by some), with fast azimuthal precession and axial mobility much lower than that of the BPW\cite{bib-THI2006,bib-YAN2010}. Both DWs have been predicted to retain their topology during motion. This makes a sharp contrast with thin strips, prone to DW transformations under both field and spin-polarized current\cite{bib-BEA2005,bib-THI2006,bib-THI2008}. The latter can be understood as all DWs share a single topology in strips\cite{bib-FRU2015b}, making transformations easier. As the existence of the BPW and TVW has been confirmed experimentally recently at rest\cite{bib-BIZ2013,bib-FRU2014}, the question arises whether the different topology indeed prevents DW transformation in reality.

In this Letter, we investigate the field-driven motion of magnetic DWs in magnetically soft nanowires. Our experiments reveal that the transformation from TVW to BPW and vice-versa may occur. We build a theoretical understanding of this topological transition, associated with the injection of a BP (or expulsion for the reverse process). Micromagnetic simulations partly confirm this qualitative description, highlighting how the precessional magnetization dynamics leads to the previously-overlooked possibility of TVW-to-BPW transformation. However, the BPW-to-TVW transformation is not found in the simulation, leaving open the question whether experiments or models should be blamed.

\section{Methods}

Starting from nanoporous alumina templates engineered with two diameter modulations along the pores, we electroplate magnetically-soft Fe$_{20}$Ni$_{80}$ and Co$_{40}$Ni$_{60}$ nanowires \cite{bib-FRU2016c,bib-FRU2015c,bib-FRU2015d}. The geometry is thus that of a cylinder with a thin section of diameter $\approx\ $\SI{140}{\nano\meter} surrounded by two wider sections of the same diameter $\approx\ $\SI{250}{\nano\meter}, all three with length~$\unit[10]{\micro\meter}$. The purpose of the modulations of diameter is to act as energy barriers around the thin section, thereby confining the DW. After fabrication, the membrane is dissolved in NaOH. After purification of the solution with water and isopropanol, the nanowires are dispersed on a silicon wafer for performing experiments on single objects. DWs are nucleated either upon demagnetization with a high magnetic field perpendicular to the wire axis, or nucleation with pulses of moderate magnetic field applied along the wire axis.

Such samples were brought to the Nanospectroscopy beamline\cite{bib-LOC2006} at synchrotron Elettra and the Circe beamline\cite{bib-ABA2015} at synchrotron Alba, to monitor DWs with shadow X-Ray Magnetic Circular Dichroism coupled to PhotoEmission Electron Microscopy (XMCD-PEEM). This technique, pioneered by Kimling \emph{et al.}\cite{bib-KIM2011b}, delivers a magnetic contrast map of the projection of magnetization along the beam having gone through the sample. This allows extracting information about magnetization in the wire core\cite{bib-FRU2014,bib-FRU2015c,bib-STR2014b}, and, combined with simulations of the shadow XMCD contrast\cite{bib-FRU2015c}, unambiguously determining the DW nature~(topology)\cite{bib-FRU2014}. In the present experiments, DWs are observed at rest, before and after the application of a quasistatic magnetic field , of duration circa $\SI{1}{\second}$, using a dedicated sample holder\cite{bib-FOE2016}.

Micromagnetic simulations were performed with our homemade finite-elements code FeeLLGood\cite{bib-ALO2014, bib-FEE}, which solves the Landau-Lifshitz-Gilbert equation\cite{bib-GIL2004}. We focused on material parameters close to those of permalloy: saturation induction $\mu_0M_\mathrm{s}=\ $\SI{1}{\tesla}, exchange stiffness $A=\SI{10}{\pico\joule\per\meter}$, and no magnetocrystalline anisotropy. This sets the dipolar-exchange length $\sqrt{2A/\muZero\Ms^2}$ at $\SI{5}{\nano\meter}$, providing the relevant length scaling, to apply the present results to any other soft magnetic material. We used $\alpha=0.05$ for the damping parameter~($\alpha=1$ was used for the relaxation of the starting configurations only), and the typical cell size was \SI{2.5}{\nano\meter}. We consider wires of finite length, however, compensate their end charges, so as to mimick infinitely-long wires. We name $z$ the direction along the wire axis, and $x$ and $y$ the orthogonal directions. While remaining in the same order of magnitude, for the sake of computational efficiency we considered  smaller diameters in the simulation~(\SI{70}{\nano\meter}) than in experiments~(\SI{140}{\nano\meter}). Comparison with existing results suggests that what matters for DW transformation is the consideration of a minimum value of diameter, however, whose precise value we did not determine precisely.

\section{Experiments}

\begin{figure}
\centering\includegraphics[width=84.941mm]{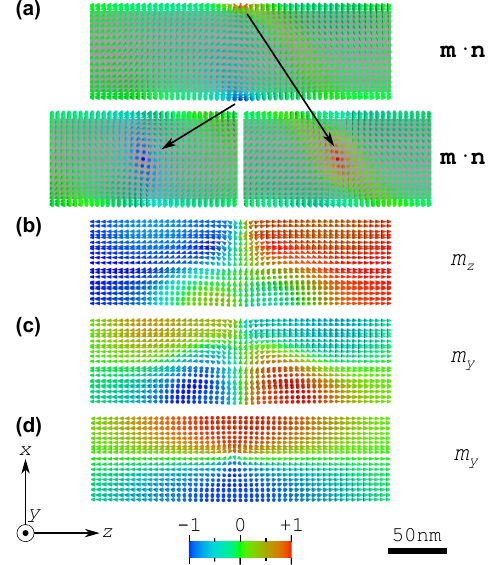}
\caption{Simulated tail-to-tail DWs in a \SI{70}{\nano\meter}-diameter wire, magnetization being represented as cones. (a) TVW surface magnetization with color coding $\vect{m}\dotproduct\vect{n}$, $\vect n$ being the outward normal to the wire surface. The gray background highlights the extent of the wire. The two bottom inset show the same rotated by $\mp\pi/2$, highlighting the vortex and antivortex of the inlet and outlet of transverse magnetization (b-c) Slice through the TVW, with $m_z$ and $m_y$ color code. (d) Slice through the BPW, with $m_y$ color code.}\label{fig_1_dws}
\end{figure}

We first recall the features of TVWs and BPWs, whose simulated configurations at rest in a \SI{70}{\nano\meter}-diameter wire are shown in \figref{fig_1_dws}. The former has dual transverse and vortex features\cite{bib-FRU2015b}. The intercept of the TVW's core (magnetized transverse to the wire axis) with the surface is highlighted by the colour map of $\vect m\dotproduct\vect n$ in \subfigref{fig_1_dws}a, where $\vect n$ is the local, outward-pointing normal to the wire surface. The core of the DW is better seen in the cross-sectional view on \subfigref{fig_1_dws}b, displaying a triangular shape reminiscent of the transverse wall in soft flat strips. The vortex feature is illustrated by the winding of magnetization around the DW core, as seen on the lower part of \subfigref{fig_1_dws}c. It has a counterpart on the opposite side~(bottom on the figure), which defines an antivortex. In contrast to this, the BPW possesses no core with transverse magnetization. It has full symmetry of rotation around the wire axis with a curling of magnetization, and a mirror symmetry perpendicular to the latter. For continuity reasons the BPW must feature a BP singularity at its center, \bracketsubfigref{fig_1_dws}{d}.

\begin{figure}
\centering\includegraphics[width=86mm]{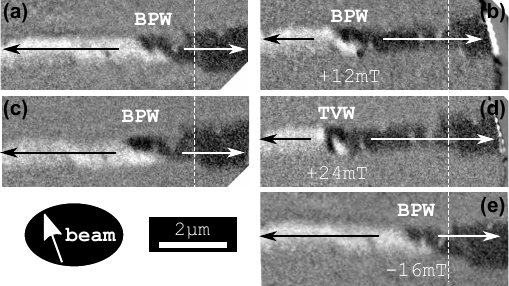}
\caption{XMCD-PEEM views of the shadow of a $\SI{140}{\nano\meter}$-diameter $\mathrm{Fe}_{20}\mathrm{Ni}_{80}$ nanowire, featuring a tail-to-tail DW in its thin section. Arrows stand for magnetization in the domains, and the vertical dotted line indicates the diameter modulation. a)-b) and c)-d) are two sequences initialized with a BPW followed by the application of a quasistatic field with strength \SI{12}{\milli\tesla} and \SI{24}{\milli\tesla}, respectively. e) follows d), after application of -\SI{16}{\milli\tesla}}\label{fig_2_xmcdpeem}
\end{figure}

\figref{fig_2_xmcdpeem} shows DWs in wires, imaged with XMCD-PEEM. Only the shadow is visible, not the wire itself, related to the focus settings. The choice of in-plane angle of incidence of the the beam, close to \ang{70} away from the wire axis~(see figure), provides magnetic contrast from both the domains and the DWs. As shown previously\cite{bib-FRU2015c}, a bipolar contrast in the shadow of the DW is indicative of curling around the axis, and is the signature of the BPW. This is the case for \subfigref{fig_2_xmcdpeem}{a-c,e}. To the contrary, the less symmetric contrast in \subfigref{fig_2_xmcdpeem}{d} is indicative of a TVW\cite{bib-FRU2015c}~(see also supplementary material).

This series of images provides two examples of DW transformation, taken among a large set of similar series. They occurred associated with DW motion between two pinning sites under a magnetic field applied along the wire axis: \subfigref{fig_2_xmcdpeem}a and \subfigref{fig_2_xmcdpeem}c are both a starting configuration with a tail-to-tail BPW. The outcome of the application of a magnetic field depends on its strength. Under $\SI{12\pm1}{\milli\tesla}$ the DW is displaced without visible changes to the configuration\bracketsubfigref{fig_2_xmcdpeem}b. To the contrary, under $\SI{24\pm1}{\milli\tesla}$ the DW is displaced and its final topology changed from BPW to TVW\cite{bib-FRU2014,bib-FRU2015c}\bracketsubfigref{fig_2_xmcdpeem}d. Note that the images all follow each other in time in \figref{fig_2_xmcdpeem}: \subfigref{fig_2_xmcdpeem}c results from re-initialization, pushing the DW backward towards the diameter modulation from \subfigref{fig_2_xmcdpeem}b, under \SI{-12}{\milli\tesla}. Repeating this experiment for the opposite (i.e. head-to-head, see supplementary) DW polarity with a sequence of opposite fields led to the same result. Transformations from a TVW to a BPW were also observed, with a similar threshold\bracketsubfigref{fig_2_xmcdpeem}e. Finally, we also investigated Co$_{40}$Ni$_{60}$ as another rather soft magnetic material, for the sake of generality. DWs may also transform from BPW to TVW and vice-versa, with a threshold of field of $\SI{9}{\milli\tesla}$, proving that the phenomenon is not limited to a specific sample or material~(Supp. \figurename~3).

\section{Theoretical description and discussion}

\begin{figure}[!t]
\centering\includegraphics[width=86mm]{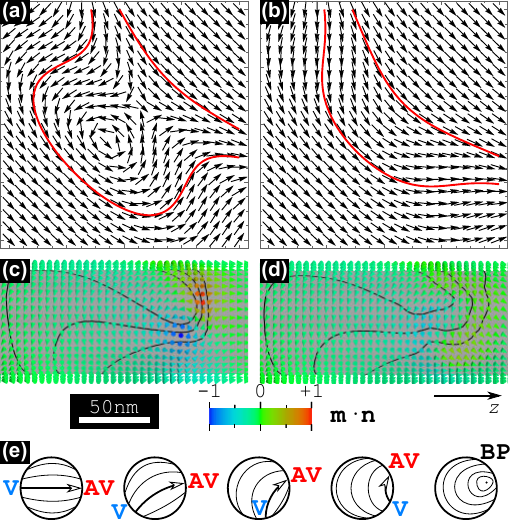}
\caption{Top row: unrolled maps of the parametrized vector field $\vect a_\mathrm{t}$ before (a) and after (b) the merging of the vortex-antivortex pair. Two field lines are displayed. Bottom: micromagnetic simulation of the transformation of a TVW under a magnetic field $\mu_0H_\mathrm{a}\!=$\SI{8.2}{\milli\tesla}, with snapshots at $t_1\!=\SI{0.45}{\nano\second}$ and $t_2\!=\SI{0.62}{\nano\second}>t_\mathrm{c}$, \ie before and after the transformation process. (e)~illustration of the process of the expulsion of the core of the TVW. V, AV and BP stand for vortex, antivortex and Bloch point, respectively.}\label{fig_3_simu}
\end{figure}

The above experimental results show that the injection or annihilation of the micromagnetic BP topological defect is possible, and does not require extremely high field amplitudes. In other words, the distinct topological natures of the BPW and the TVW do not prevent these DWs to transform into one another. However, the transformation process, of precessional nature\cite{bib-VAN2006}, cannot be grasped by our static imaging. Therefore, we first turn to a simplified theoretical description to understand the path followed during the transformation of texture. We define a unit vector field $\vect a_\mathrm{t}$ standing for magnetization at the surface of a cylinder and tangential to the local surface. It is chosen in such a way that it features the aforementioned vortex-antivortex pair, and respects the boundary conditions with domains of opposite orientation at both ends. $\vect a$ is defined from the unit vector orthogonal to the gradient of a third-order polynomial defined on a planar surface. $\vect a_\mathrm{t}$ is thus always tangent to the cylinder~(see Supplementary). One of the polynomial's coefficients is parametrized with a pseudo-time $\tau$ so that at $\tau=0$, the polynomial has a local minimum corresponding to the vortex, and a saddle point corresponding to the antivortex. As $\tau$ increases, the polynomial's local minimum and saddle point are brought closer\bracketsubfigref{fig_3_simu}{ab}, until they merge  at a critical time $\tau_\mathrm{c}<1$ into an inflexion point. Later on, for $\tau_c<\tau<1$, the polynomial no longer has any extremum nor inflexion point\subfigref{fig_3_simu}{b}. The time line can be reversed to equally describe the transformation of a BPW into a TVW.

The relevance of this construction stems from the identical boundary conditions at any time, the initial presence of a vortex-antivortex pair, and also from the $\mathcal{S}^2$ winding number $N_\mathrm{p}(\tau)$ associated to the parametric vector field. Indeed, $N_\mathrm{p}(\tau<\tau_c)=0$, while $N_\mathrm{p}(\tau>\tau_c)=1$, which indicates that any normalized continuation of $\vect a_\mathrm{t}$ into the cylinder's interior \emph{must} contain a singular point. Therefore, we have built with simple tools a unit vector field for a possible mechanism for the TVW-to-BPW transformation: the vortex-antivortex pair is annihilated on the wire surface at $\tau_\mathrm{c}<1$, associated with the injection of a BP into the volume. The abrupt character of this injection is highlighted on \subfigref{fig_3_simu}{a,b}, with the vector field $\vect a$ displayed before and after $\tau_c$. A pair of field lines are displayed. They encompass the vortex-antivortex pair and hence a $2\pi$ rotation of $\vect a_\mathrm{t}$ before $\tau_\mathrm{c}$, while later the region they encompass is narrower, with overall smoother variations of $\vect a_\mathrm{t}$. The disappearance of the \ang{360}-DW-like configuration highlights the discontinuous transformation undergone by the vector field.

The mechanism described above is plausible from the mathematical point of view, but its physical validity must be checked with micromagnetic simulations. To that end, we start from a tail-to-tail TVW configuration in a \SI{70}{\nano\meter}-diameter and \SI{1}{\micro\meter}-long FeNi wire relaxed under zero magnetic field. We then apply a constant field in a step-wise manner at $t=0$. We evidence the existence of a threshold field in the simulations, consistent with the experiments. In the low-field regime the DW precesses around the wire axis while slowly moving forward, soon reaching a steady-state regime in a rotating frame. This is consistent with prior knowledge about TVWs\cite{bib-HER2001,bib-FOR2002,bib-THI2006}. We detail below the high-field regime, with the selected case $H=\SI{8.2}{\milli\tesla}$ slightly above the threshold. After $t_1$=\SI{0.45}{\nano\second} the vortex and antivortex are no longer diametrically opposed but rather close to one another, as shown by the neighbouring extrema of opposite signs in the colour map of $\vect m\dotproduct\vect n$\bracketsubfigref{fig_3_simu}c. This dynamical effect is not surprising, as two aspects break the symmetry between the vortex and antivortex: 1/~owing to the triangular shape of the TVW\bracketsubfigref{fig_1_dws}{b-c}, they occupy non-equal volumes within the DW. If one applies handwavingly the one-dimensional model to surface magnetization, the wall width is different at both sites, so that a given torque arising from the applied field translates into different velocities 2/~The gyrovectors of the vortex and antivortex differ, contributing to a difference in both the longitudinal and azimuthal motion.

We write $N_\mu$ the winding number in the simulations. While $N_\mu(t_1)\approx0$, the calculated surface isolines of $m_z=-0.8$, $0$ and $0.8$ are quite pinched. At $t_2\!=\SI{0.62}{\nano\second}$, these curves have become more separated and smooth, and visually the vortex and antivortex have both disappeared\bracketsubfigref{fig_3_simu}d. This suggests the injection of a Bloch point resulting from the recombination into the volume of the vortex-antivortex annihilation at the surface, which is formally confirmed as $N_\mu(t_2)\approx1$.

Another view is the following: the line of magnetization linking the surface vortex and antivortex through the wire, initially the straight transverse component, becomes an arc getting shorter and shorter over time, until being completely expelled from the wire at~$t_\mathrm{c}$\subfigref{fig_3_simu}{e}.  Note that the micromagnetic time evolution of surface magnetization is consistent with the parametrized model\bracketsubfigref{fig_3_simu}{a,b}.

\begin{figure}[t]
\centering\includegraphics[width=81.737mm]{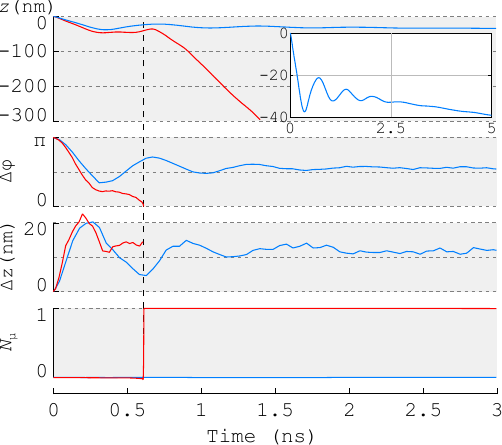}
\caption{Numerical micromagnetic indicators versus time for DW motion below~(blue, $\unit[6]{\milli\tesla}$) and above~(red, $\unit[8.2]{\milli\tesla}$) the threshold field~$\Hc$: winding number $N_\mu$, azimuthal and axial separation of vortex versus antivortex $\Delta\varphi$ and $\Delta z$, position~$z$. The vertical line indicates the time of change of topology}\label{fig_4_simu_curves}
\end{figure}

With a view to refine the microscopic process underlying the transformation, we monitor the magnetization texture over time with a series of numerical indicators\bracketsubfigref{fig_4_simu_curves}. First, monitoring $N_\mu(t)$ confirms an abrupt jump consistent with the nucleation of a BP and the change of topology, from which we define the critical time $t_\mathrm{c}\approx\SI{0.61}{\nano\second}$. To confirm the nucleation process we need to monitor the vortex and antivortex with a more robust tool than the normal surface magnetization $\vect m\dotproduct\vect n$.  Topology is well-suited to track localized singular objects, however, vortices and antivortices are not singularities for magnetization on the unit sphere. This is why we consider $\vect m_t$, the normalized 2D projection of surface magnetization onto the local tangent plane. The vortex and antivortex \emph{are} now singularities for this vector field defined on the unit circle~$\mathcal{S}^1$. Thus, along any contour $\Gamma$ (on the wire surface) surrounding the vortex or antivortex, $\vect m_t$ wraps exactly once around the unit circle. In other terms, the corresponding $\mathcal{S}^1$ winding number is $N_\Gamma\pm1$. A bracketing algorithm can then be used to track the position of both surface objects over time, from which we compute the difference in angular position $\Delta\varphi(t)=\varphi_\mathrm{AV}(t)-\varphi_\mathrm{V}(t)$ and in abscissa along the wire axis $\Delta z(t)=z_\mathrm{AV}(t)-z_\mathrm{V}(t)$. Starting from $\Delta\varphi(0)=\pi$ and $\Delta z(0)=0$ for the initial TVW, $\Delta\varphi$ decreases monotonuously, while $\Delta z(t)$ first increases before decreasing again. Ultimately, both quantities become undefined at $t_\mathrm{c}$, providing a firm ground to the previous discussion. Finally, the DW velocity sharply increases soon after the transformation~[see $z(t)$ on \figref{fig_4_simu_curves}], in agreement with existing knowledge regarding BPWs versus TVWs\cite{bib-HER2004a,bib-THI2006}. \figref{fig_4_simu_curves} also shows the time evolution of those numbers below the field threshold for the transformation. The initial trend is similar, the vortex moving faster then the antivortex axially, and rotating slower. However, $\Delta\varphi$ and $\Delta_z$ eventually reach a plateau instead of vanishing. The probable reason is that the field-driven torque is too weak to overcome the cost in exchange when the vortex and antivortex draw nearer. This also probably explains why the DW transformation in wires of constant diameter had been overlooked in simulations so far: significantly smaller diameters were considered, associated with a larger exchange cost to bend the DW core and join vortex and antivortex, preventing the transformation.

 We now come back to the experiments, in light of the above, and proceed to the final discussion. Experiments and theory agree on the existence of a threshold field for the TVW-to-BPW transformation, and also on its value around $\unit[10]{\milli\tesla}$. The quantitative agreement is surprisingly good, given the possible interplay of material imperfections and thermal activation in experiments, and the difficulty to describe accurately sharply-varying magnetization textures and BPs in micromagnetics\cite{bib-THI2003,bib-ELI2011,bib-LEB2012,bib-HER2014b}. To the contrary, we observe the reverse process only in the experiments~(BPW to TVW), again with the existence of a threshold~(however, with typically $\unit[50]{\%}$ higher value), but not in the simulations. Instead, a steady-state motion of the BPW occurs. In some cases we evidenced spiraling instabilities around the wire axis, encountered when the mesh is not fine enough or time step too large. However, others also reported spiralling instabilities and emission of magnetic drops at high field with a finer code combining micromagnetics and atomistic modelling, although, still with no transformation\cite{bib-HER2015}. At this point it remains an open question whether it is imperfections in the experiments or in theory, which are to be blamed. This discrepancy in describing a micromagnetic phenomenon is fascinating, while numerical micromagnetics has become ripe in accuracy and power to be considered as a key predictive tool.

\section{Conclusion}

We observed the topological transformation of domain walls in cylindrical nanowires upon motion under magnetic field, from the transverse-vortex type~(TVW) to the Bloch point type~(BPW) and the reverse, with a threshold field in the range of~$\unit[10]{mT}$. This is fundamentally different from the Walker breakdown in flat strips, which preserves the topology. Micromagnetic simulations reproduce quantitatively the TVW-to-BPW transformation, involving the nucleation of a Bloch point at the wire surface, however, not the BPW-to-TVW transformation, which may arise from imperfections in the experiment, or lacking ingredients in the simulation. This unique micromagnetic case escaping our understanding raises exciting challenges in material science and numerical computation to solve the discrepancy. Also, it shows that care should be taken when searching for the specific features predicted for the motion of BPWs, namely absence of breakdown field\cite{bib-THI2006,bib-HER2015} and the possible spin-Cherenkov effect\cite{bib-YAN2011b}.

\begin{acknowledgments}
This project has received funding from the European Union Seventh Framework Programme (FP7/2007-2013) under grant agreement n\textsuperscript{$\circ$} 309589 (M3d). M.S. acknowledges grant from the Laboratoire d'excellence LANEF in Grenoble (ANR-10-LABX-51-01). The PEEM experiments performed at ALBA Synchrotron involve the collaboration of ALBA staff, with special thanks to Jordi Prat. We thank Benjamin Canals from Institut N\'{e}el and Michel Hehn from Institut Jean Lamour for fruitful discussions.
\end{acknowledgments}

\bibliographystyle{apsrev}

\end{document}